\documentclass{aa}  
\usepackage{natbib}
\bibpunct{(}{)}{;}{a}{}{,} 
\usepackage{float}
\usepackage{times}

\usepackage{graphicx}
\usepackage{capt-of}
\usepackage{booktabs} 

\usepackage{txfonts}

\usepackage{multirow}
\usepackage{stfloats}

\usepackage{xcolor}
\usepackage[normalem]{ulem}

\begin{document}

   \title{WLM: Dynamics of an isolated Dwarf Irregular Galaxy Under Ram Pressure in the Local Group}

   \author{Neel Kolhe\inst{1}
          \and 
          Francois Hammer\inst{1}
          \and
          Yanbin Yang\inst{1}    
          \and
          Brenda Namumba\inst{2}
          \and
          Laurent Chemin\inst{3,4}
          \and
          Philippe Amram \inst{5}
           \and
          Roger Ianjamasimanana\inst{2}
          \and
          Claude Carignan\inst{6,7,8}
          }

    \institute{
      LIRA, Observatoire de Paris, Universite PSL, CNRS, Place Jules Janssen, 92195 Meudon, France\\
      \email{neel.kolhe@obspm.fr}
      \and
      Instituto de Astrof\'isica de Andaluc\'ia (CSIC), Glorieta de la Astronom\'ia s/n, 18008 Granada, Spain
      \and
        Université de Strasbourg, CNRS, Observatoire astronomique de Strasbourg, UMR 7550,
        67000 Strasbourg, France
       \and
       Instituto de Astrofísica, Departamento de Ciencias Físicas, Universidad Andrés Bello, Fernández Concha 700,
        \and
        Aix Marseille Univ, CNRS, CNES, LAM, Marseille, France
       \and
       Department of Astronomy, University of Cape Town, Private Bag X3, Rondebosch 7701,South Africa
       \and
       Département de physique, Université de Montréal, Complexe des sciences MIL, 1375 Avenue
       Thérèse-Lavoie-Roux, Montréal, QC, Canada H2V 0B3
       \and
       Observatoire d’Astrophysique de l’Université Ouaga I Pr Joseph Ki-Zerbo (ODAUO), BP 7021, Ouaga 03, Burkina Faso
        }

  \abstract
  { The Wolf--Lundmark--Melotte (WLM) galaxy is an archetypal dwarf irregular galaxy that has not experienced interactions with major Local Group galaxies within the past 8 Gyr. It has recently been shown that WLM is losing its gas due to ram pressure forces exerted by the surrounding intergalactic medium. In this work, we explored how ram pressure can also affect the WLM gas kinematics, and we show that its dynamics is especially perturbed at its outskirts, which explains the asymmetric rotation between the approaching and receding sides. Moreover, we have been able to decompose WLM into two main components, a compact one with a solid-body rotation that resembles a bar-like structure, and a more extended one with a characteristic double-horn profile suggesting an edge-on disc. The former is relatively unaffected by ram pressure while the dynamics of the latter is considerably affected. This study shows that mass estimates of a dwarf galaxy like WLM should involve a full modelling of its dynamical components, especially its asymmetric rotation curve.}


\maketitle
%

\section{Introduction}

The Wolf–Lundmark–Melotte (WLM) galaxy is an isolated gas-rich dwarf irregular galaxy in the Local Group (LG) roughly 900 kpc away from the Milky Way and M31 (McConnachie, 2012).
WLM is one of the few dwarf galaxies within 1 Mpc that have not experienced interactions with the Milky Way or with M31 \citep{higgs_solo_2021}.
Owing to this spatial isolation throughout its evolutionary history, WLM has been a subject of extensive studies and has been characterised in the literature as a 'prototypical irregular dwarf'. In the hierarchical merger scenario of galaxy formation, dwarfs like WLM  have merged to form larger galaxies like the Milky Way and M31. Hence studying an isolated dwarf like WLM gives us an insight into the properties of these progenitor dwarfs. 
Its high inclination has also made it a good candidate for dynamical studies \citep{hunter_little_2012}. However, the $70^\circ$ inclination poses a challenge to determining features within the thick disc, which has been described as an oblate spheroid. 

WLM therefore has a rich history of being observed by flagship telescopes across frequencies. 
Recent combined JWST and Hubble Space Telescope studies have probed its star formation history \citep{mcquinn_jwst_2024,albers_star_2019}, revealing steady star formation for the past 7 billion years owing to its gas-rich nature. Molecular clouds contributing to this star formation traced by CO cores have also been extensively studied by the Atacama Large Millimeter/submillimeter Array (ALMA) \citep{archer_environments_2022} and in UV by the Ultraviolet
Imaging Telescope aboard ASTROSAT \citep{mondal_uvit_2018}. 
HI Radio observations have been performed with the Very Large Array (VLA), Green Bank Telescope (GBT) and MeerKAT telescopes. 
They have provided us with an exquisite view into the neutral gas dynamics of WLM. For over a decade it has been noted in the literature, first by \citet{kepley_high-resolution_2007}, that WLM shows a lopsided rotation curve on the approaching side of the galaxy. This triggered work to explain the asymmetry by considering a perturbed halo \citep{khademi_kinematical_2021}, but the absence of past interactions disfavours this hypothesis.

\citet{yang_evidence_2022} found evidence of ram pressure stripping in the galaxy with the discovery of four trailing clouds detected by MeerKAT in the direction opposite to the galaxy's proper motion which was obtained from Gaia Data Release 3 by \citet{battaglia_gaia_2022}.
They were also able to show that the clouds had no stars associated with them, confirming that the gas has been stripped through ram pressure.
\citet{yang_evidence_2022} were able to replicate the orientation and mass of these stripped clouds using hydrodynamical simulations. They considered multiple models of WLM through various densities of the intergalactic medium (IGM) and assuming  varying WLM velocities with respect to the IGM, to constrain its density in the surroundings of the LG. \cite{mcconnachie_ram_2007} reported a similar occurrence of gas loss due to ram pressure induced by the IGM within the LG for the Pegasus dwarf galaxy (DDO 216), which is also an isolated body.

\begin{table}[h!]
\caption{Relevant Properties of WLM}
\label{tab:Properties}
\centering
\renewcommand{\arraystretch}{1.3}
\setlength{\tabcolsep}{3pt}
\begin{tabular}{lllll}
\hline\hline
Parameters & Symbol & Value & Units & Ref. \\
\hline
Optical Center & RA & $\mathrm{00^h\,01^m\,58.2^s}$ & - & 1 \\
-         & Dec & $\mathrm{-15^\circ\,27'\,39'}$ & - & 1 \\
Dynamical Center & RA & $\mathrm{00^h\,01^m\,57.9^s}$ & - & 2 \\
-              & Dec & $\mathrm{-15^\circ\,27'\,12.26'}$ & - & 2 \\
Systemic Velocity  & $V_{sys}$ & -126 & km s$^{-1}$&  2\\
Distance  & d & 0.934 & Mpc&  \\
Inclination  & Inc & 75.11 & $\circ$&  2\\
Position Angle  & PA & 175.5 & $\circ$&  2\\
Scale Height  & H & 0.4 & kpc&  3\\
Stellar Mass  & $M_\star$ & $4.3 \times 10^7$ & $M_\odot$&  4\\
HI Mass (MK16)  & $M_{HI}$ & $6.1 \times 10^7$ & $M_\odot$&  2\\
HI Mass (MK64)  & $M_{HI}$ & $6.5 \times 10^7$ & $M_\odot$&  2\\
Comp 1 HI Mass & $M_{HI}$ & $3.4 \times 10^7$ & $M_\odot$&  2\\
Comp 2 HI Mass & $M_{HI}$ & $3.0 \times 10^7$ & $M_\odot$&  2\\
\hline
\end{tabular}
\label{table1}
\tablefoot{1-\citet{higgs_solo_2021}.
2-This work.
3-\citet{ianjamasimanana_meerkat-16_2020}. 4-\citet{mcconnachie_observed_2012}.
Dynamical Center, Systemic Velocity, Inclination and Position Angle are derived from modelling presented in Section 4. }
\end{table}

A large reservoir of baryonic matter in the Universe is expected to be present in IGM filaments. Observational regimes ranging from fast-radio-bursts to Lyman-alpha-based methods have been used to place constraints on the cosmic IGM density \citep{Connor2025,jin_nearly_2023}. 
An intra-cluster medium of density $10^{-2}$ - $10^{-3}$ $atoms/cm^3$  in large clusters has been studied and quantified using X-ray telescopes \citep{li_gas_2024}.
However, it is much sparser in less massive systems such as the LG, making it very hard to detect. 

HI gas discs extend much farther beyond the stellar discs, especially for dwarf irregulars, and therefore, can be used to probe the total mass content of a galaxy to larger radii \citep{Carignan_DDO_154,de_blok_high-resolution_2008,patra_figgs2_2016}, under the assumption of dynamical equilibrium.  In the case of WLM, an extreme dark matter (DM)-to-baryon ratio of 90:1 has been reported \citep{read_stellar_2017}.
However, since the pioneering Gunn \& Gott (1972) study, it is well established that the motion of galaxies through intra-group or intra-cluster medium perturbs their kinematics, particularly at the outer edges of their gaseous discs. Such interactions can drive the gas out of equilibrium, thereby biasing rotation curves and leading to systematic errors in the estimation of galactic masses.
Therefore, having a thorough understanding of the DM halo extension and the effect of the IGM on gas-rich bodies is necessary to estimate the total mass content of dwarf irregulars. 

There is growing evidence that ram pressure stripping due to the circumgalactic medium of the Milky Way plays a major role in removing gas and transforming the dynamics of the Milky Way dwarfs \citep{hammer_accretion_2023,wang_accretion_2023}. Gas loss in WLM could also affect its dynamical evolution. The steady stripping of gas could also contribute to the two broad historical sub-classifications of isolated irregular dwarfs, those being gas-rich and the other being mostly devoid of gas. Therefore, the goal of this paper is to use insights from state-of-art high-velocity-resolution  21cm radio observations of WLM from the MeerKAT telescope to study the kinematics of WLM and the effect of the IGM on gas dynamics.

In section 2 we revisit the \citet{ianjamasimanana_meerkat-16_2020} MeerKAT 16-dish (MeerKAT-16 hereafter) data to characterize the stripping gas more robustly and elaborate on the data reduction of the new MeerKAT 64-dish (MeerKAT-64) observations of WLM. In section 3, we look at the gas distribution in WLM and present a two-component Sérsic fit. In section 4 we derive the rotation curve for WLM and discuss its dynamical properties. Section 5 elaborates on the kinematical decomposition that we performed on the MeerKAT-64 data, and in section 6 we discuss the consequences of our findings.

\section{MeerKAT observations}
\subsection{Revisiting the MeerKAT 16 dish observations}

\cite{yang_evidence_2022} used the observations from \cite{ianjamasimanana_meerkat-16_2020}, where they generated a mask by blanking pixels with values less than 3 times the RMS noise in the cube and used the MAFIA task in the MIRIAD software suite, leading them to detect the four stripped clouds. 
Motivated by trying to make a more robust detection of the stripped gas, we re-analysed the data cube with the SoFiA-2 HI source finding tool. SoFiA-2 is optimised for HI surveys and can take into account noise variations and it handles artefacts more reliably. We used the Smooth + Clip finder which iteratively smooths the cube with a defined set of spatial and spectral smoothing kernels, measuring the noise level at each smoothing scale and adding all pixels with an absolute flux density above a relative threshold \citep{serra_sofia_2015,westmeier_span_2021}.
We set the source finding threshold to be 4 times the RMS and used spatial smoothing kernels of sizes 3, 6, 9 , 12, 15, and 18 pixels; as 3 pixels is just under the beam size and the smallest of the clouds detected by \citet{yang_evidence_2022} is roughly 18 pixels wide.

\begin{figure}[h!]
  \resizebox{\hsize}{!}{\includegraphics{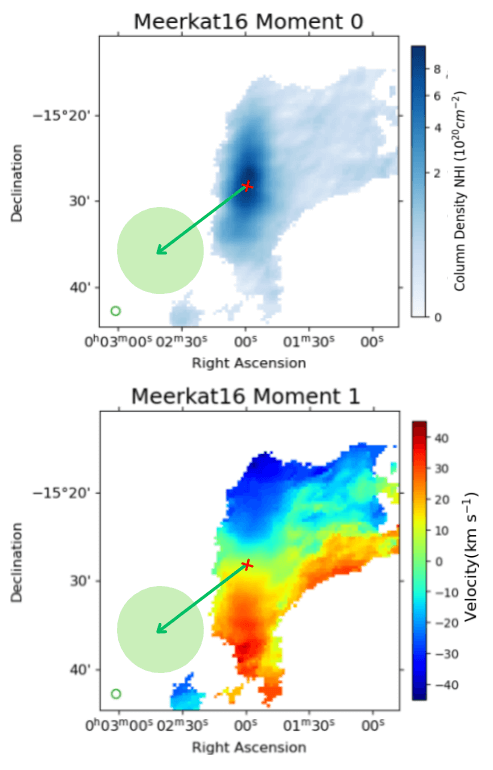}}
  \caption{\ {Our new masked MeerKAT-16 moment-0 \textit{(top)} and moment-1 maps \textit{bottom} of WLM. Both panels reveal the stripped HI tail. The green arrow indicates proper motion and the transparent circle shows the 1$\sigma$ uncertainty as presented in \citet{battaglia_gaia_2022} and \citet{yang_evidence_2022}.} The green circle at the bottom left indicates the beam size. }
  \label{mk16}
\end{figure}
With the new SoFiA-2 mask applied to the data cube, we derived the moment maps presented in Fig.~\ref{mk16}. Instead of four isolated stripped clouds, we detect a continuous tail of stripped gas, akin to jellyfish galaxies \citep{roberts_lotss_2021}.
For the main body of the galaxy, in the same region defined by \citet{yang_evidence_2022}, we recover their total HI mass value of $6.1 \times 10^7$\(M_\odot\). However, \citet{yang_evidence_2022} have found a mass of $6.1 \times 10^6$\(M_\odot\) for the stripped gas, while we recover an improved $7.7 \times 10^6$\(M_\odot\). 

Because we now see the tail of the stripped gas attached to the main body, we can also determine where the gas has been stripped from as we see a direct connection to the main body in the velocity field. The velocities distributions across the main body are clearly against the proper motion vector.

\subsection{MeerKAT 64 Observation}

\begin{table}[ht]
\caption{H\,\textsc{i} Observations and Imaging Parameters for the WLM Galaxy}
\label{tab:obsparams}
\centering
\begin{tabular}{l r}
\toprule
\textbf{Parameter} & \textbf{Value} \\
\midrule
Project code               & SCI-20220822-BN-01 \\
Observing dates            & 2023 Jul 2--4 (UTC) \\
Array configuration        & MeerKAT (61 dishes) \\
Mosaic pointings           & 11 \\
Band                       & L band (856 MHz) \\
Spectral resolution  & 0.7 km\,s$^{-1}$ \\
uv-taper (final cubes)     & 50$''$ \\
Weighting                  & Robust 0.0 \\
\midrule
\multicolumn{2}{l}{\textit{Mosaic parameters}} \\
\midrule
Synthesized beam           & $64'' \times 60''$ \\
RMS noise (single pointing) & 0.0034 Jy\,beam$^{-1}$ \\
RMS noise (mosaic)         & 0.0018 Jy\,beam$^{-1}$ \\
Beam area scatter          & 3.78\% \\
Flux density uncertainty   & $\sim$7.6\% \\
3$\sigma$ $N_{\mathrm{H\,\textsc{i}}}$ sensitivity & $1.1 \times 10^{18}$ cm$^{-2}$ \\
\bottomrule
\end{tabular}
\end{table}
Neutral hydrogen (H\,\textsc{i}) observations of the WLM galaxy were obtained with the MeerKAT radio telescope \citep{Jonas:2018Jr} as part of project \textsc{SCI-20220822-BN-01}, over two observing epochs on 2023 July 2--3 and July 3--4 (UTC). Observations were conducted in L-band 32k mode, which provides 32,768 frequency channels across an 856~MHz total bandwidth. For the purposes of H\,\textsc{i} spectral line imaging, a subset of 701 channels centred on the rest frequency of the H\,\textsc{i} line (1420.4058~MHz) was extracted, corresponding to a 2.289~MHz bandwidth. This yields a velocity coverage of $\sim483$~km\,s$^{-1}$ at a spectral resolution of 0.7~km\,s$^{-1}$ per channel.

To map the H\,\textsc{i} distribution in and around WLM, we used an 11-pointing mosaic covering a rectangular area of $3^{\circ} ~ \times ~ 2.35^{\circ}$. The mosaic was arranged asymmetrically and extended towards the north-west, specifically designed to encompass several H\,\textsc{i} clouds previously detected in lower-resolution MeerKAT-16 observations. This layout provided complete coverage of both the central disc and the surrounding low-column-density H\,\textsc{i} environment. Each pointing was observed for a total integration time of approximately one hour. The primary flux and bandpass calibrator J1939--6342 was observed every 3 hours, while the complex gain (phase) calibrator J0025--2602 was observed approximately every 30 minutes.

The data were processed on a virtual machine of the Spanish Prototype of the Square Kilometre Array (SKA) Regional Center(espSRC)\citep{espsrc}. Both observing epochs were calibrated using the Containerised Automated Radio Astronomy Calibration (CARACal) pipeline \citep{Jozsa2020}. CARACal provides an environment to carry out standard calibration and reduction steps, including radio frequency interference flagging, cross-calibration, self-calibration, continuum subtraction, and imaging. 

H\,\textsc{i} spectral line imaging was performed individually for each pointing using WSClean \citep{2014MNRAS.444..606O} as part of the CARACal pipeline. We followed a two-step iterative strategy to produce the final image cubes. In the first step, a low-resolution cube was created using a Gaussian uv-taper of 90$^{\prime\prime}$ and brings Briggs robust weighting of 1.5, to enhance sensitivity to extended low-surface-brightness emission. These cubes were cleaned using WSClean’s automatic masking at a 5$\sigma$ threshold, and the resulting masks were refined using SoFiA-2 \citep{westmeier_span_2021} to isolate real emission structures. In the second step, the refined masks were applied to produce final cubes at higher resolution using a uv-taper of 50$^{\prime\prime}$. These final cubes were cleaned down to 0.5$\sigma$ per channel, ensuring both completeness of faint emission and suppression of sidelobe artefacts.

The final data product is a linear mosaic composed of 11 pointings, each with a channel width of 0.7\,km\,s$^{-1}$ and a rms noise per channel of approximately 0.0034\,Jy\,beam$^{-1}$, measured before primary beam correction. The mosaic was produced using CARACal, with primary beam correction applied during the mosaicing process. The resulting synthesised beam size is 64$''$~$\times$~60$''$, and the average rms noise in line-free regions of the central, primary-beam-corrected mosaic area is approximately 0.0018\,Jy\,beam$^{-1}$ per 0.7\,km\,s$^{-1}$ channel. This imaging configuration was chosen to match the synthesised beam size of prior MeerKAT-16 observations, enabling direct comparisons across datasets and resolutions.

To account for beam variations among individual pointings across the mosaic, we estimated a relative scatter in beam area of 3.78\%, which leads to a maximum flux density uncertainty of approximately 7.6\%. Considering this uncertainty and the achieved sensitivity, the mosaic provides a 3$\sigma$ H\,\textsc{i} column density sensitivity of approximately $1.1 \times 10^{18}$\,cm$^{-2}$ over a 20\,km\,s$^{-1}$ linewidth in the central region. The observation setup and relevant parameters are summarised in Table~\ref{tab:obsparams}.

\begin{figure*}[h!]
  \resizebox{\hsize}{!}{\includegraphics{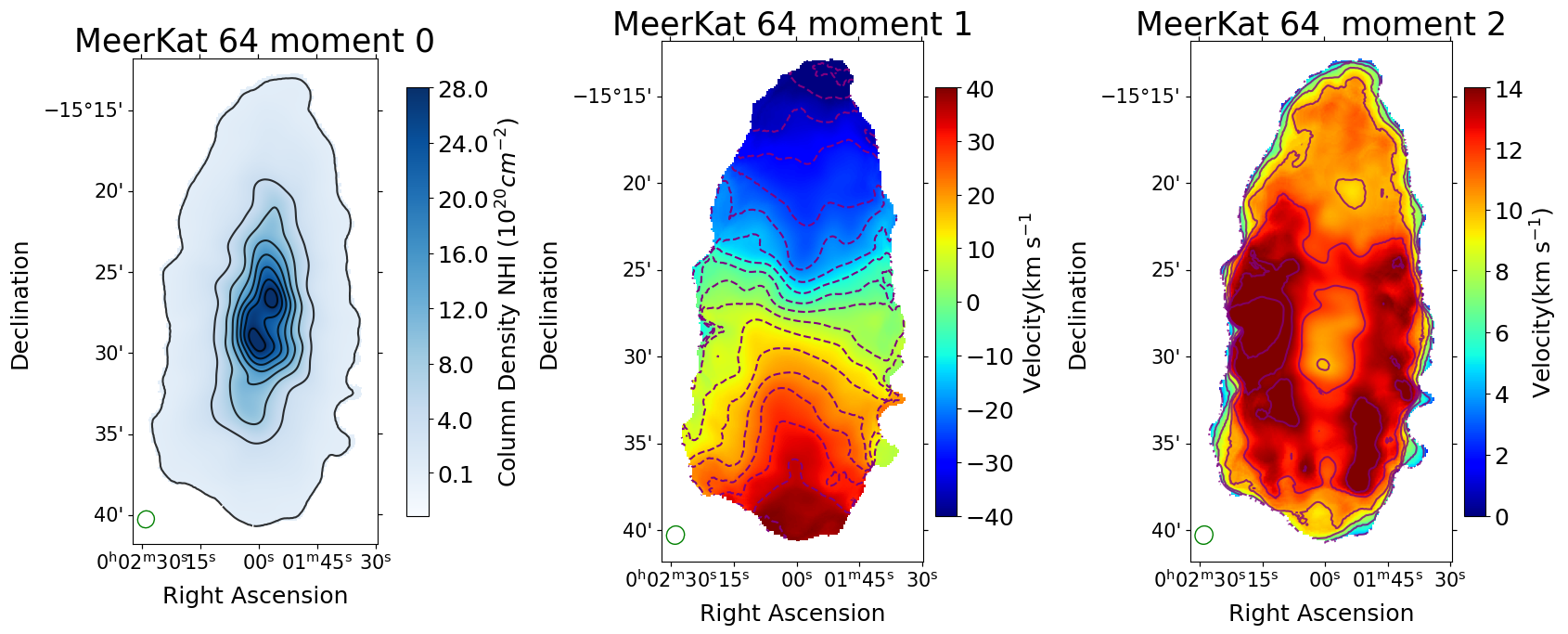}}
  \caption{\ {HI high-resolution moment maps of the MeerKAT-64 observation of WLM. \textit{Left:} Our new MeerKAT-64 moment-0 map. All the figures using the MeerKAT-64 data in this paper are masked by this map. The contours of the moment-0 map correspond to the ticks of the colour bar. \textit{Centre:}The rest frame  moment-1  velocities. \textit{Right:}Moment-2 dispersion map corrected for instrumental broadening. The contours of the moment-1 map correspond to a velocity range between -40 and 40 km s$^{-1}$ with a spacing of 5 km s$^{-1}$ and for the moment-2 map they correspond to 6,9 and 12 km s$^{-1}$, respectively  The green circle shows the beamsize. }}
  \label{mk64}
\end{figure*}

The data were imaged with using natural weighting, several robust values, uv-tapers, and multi-scale deconvolution before settling on the above. The stripped gas detected in the earlier MeerKAT-16 observations was enabled by the compact 16-antenna configuration, which provided dense sampling at short baselines. In the 61-dish configuration used for our mosaic, the relative number of short spacings is significantly reduced, while the number of long baselines increases. This improves angular resolution but reduces sensitivity to large-scale, low-surface-brightness emission. We do not detect extended stripped faint structures with the 61-dish configuration, notably even features from the Magellanic Stream that appear clearly in single dish Parkes Galactic All Sky Survey (GASS) observations \citep{GASS1} in the same velocity range of the mosaic field remain undetected, indicating that the limitation is intrinsic to the UV coverage rather than the imaging strategy. A full recovery of the diffuse tail would require combining our data with single-dish observations or acquiring targeted compact-configuration interferometric data. As this lies outside the scope of the present work, we plan to address the extended cloud in a dedicated follow-up study.
While the reduced short-baseline sensitivity prevents recovery of these structures, including the diffuse tail, the 61-dish mosaic provides a high spectral resolution view of WLM’s main body and the spatial resolution necessary for a detailed kinematic analysis.
The final mosaic is once again processed with SoFiA-2 to generate a mask with the smooth+clip finder spatial kernels set to 0,3,6 and 0, 3, 7, 15 
for the spectral axis with the cubelets feature turned on, which gives us individual data cubes for each continuous source. A threshold of 5 $\times$ RMS was set for detection. The cubelet that detected the main body of the galaxy was chosen as the data source, which was used in the study. 
The generation of moment maps, column densities, flux and HI mass reported in Fig.~ {\ref{mk64}} and elsewhere in this study follow the same procedure and formulae reported in   \citet{ianjamasimanana_meerkat-16_2020}. 

We recover a total flux density of 316 Jy km\,s$^{-1}$ from our new observations, which corresponds to a total HI mass of $6.5 \times 10^7$  $M_\odot$ at a distance of 0.934 Mpc given in \citet{higgs_solo_2021}.

\begin{figure}
  \resizebox{\hsize}{!}{\includegraphics{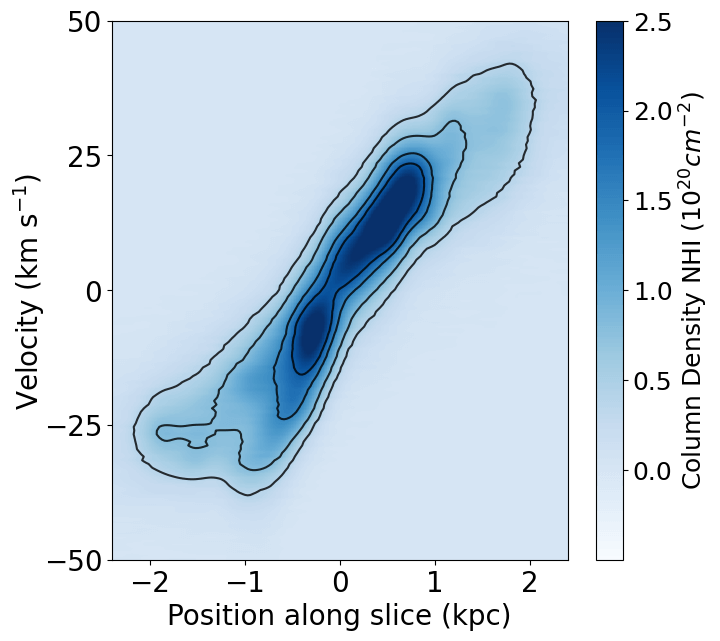}}
  \caption{Global PV diagram of WLM taken along the PA reported in Table 1. The contours correspond to the ticks on the colour bar. }
  \label{PV}
\end{figure}

\section{HI distribution in WLM}

\begin{figure}[h!]
  \resizebox{\hsize}{!}{\includegraphics{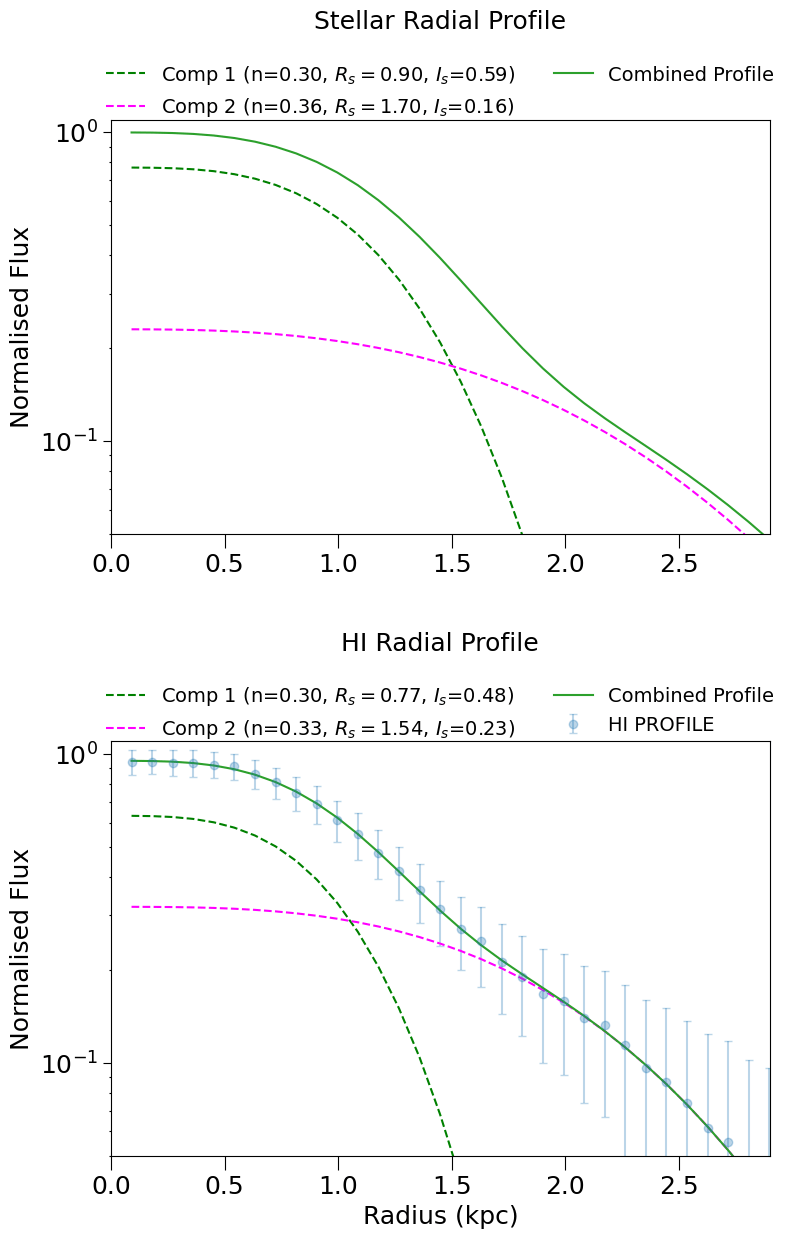}}
  \caption{\ { Top: Reproduction of the two Sérsic components of the stellar profile of WLM from \citet{higgs_solo_2021}. Bottom: Two Sérsic components of the HI profile.}}
  \label{profile}
\end{figure}

\begin{figure}[h!]
  \resizebox{\hsize}{!}{\includegraphics{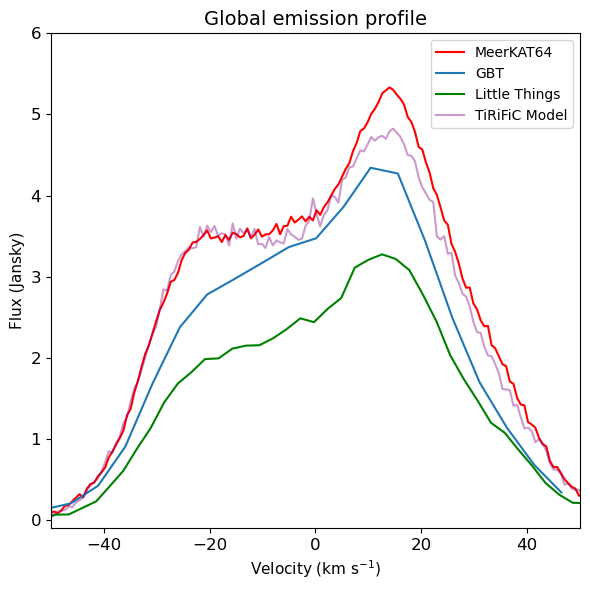}}
  \caption{\ { Emission profile of the whole body of WLM from three major HI observations. The GBT observations are from the same data cube used in \citet{ianjamasimanana_meerkat-16_2020}, and little things VLA data are from \citet{hunter_little_2012}. All observations were masked by SoFiA-2 and a common velocity range was set. Velocities are in the rest frame }.}
  \label{globalprofile}
\end{figure}

A visual inspection of our high-resolution moment-0 map in Fig.~\ref{mk64} shows us that WLM has a dense central structure that tapers into the rest of the gaseous disc. This is also seen in the corresponding position-velocity (PV) diagram in Fig.~\ref{PV}, where the dense gas core is seen along the steep velocity gradient at the centre of the galaxy.

 \citet{higgs_solo_2021} found that the stellar radial profile of WLM in their g- and i-band survey is better fitted by 2 distinct morphological components instead of a single one, both components being described by a compact Sérsic profile in the inner region and a shallower Sérsic profile in the outer region.
They tested whether a single component describes the profile better than two components by using the Akaike information criteria (AIC) and the Baysian Information Criteria (BIC). In their study for WLM a two-component model is almost 600 times more likely than the single component model, significantly favouring the two-component case.
We repeated this test for the 1D gas profile of WLM.

  Classical ellipse fitting to obtain a 1D HI flux profile may not work well for a galaxy at a very high inclination such as WLM. Therefore, we used two separate methods to  extract the 1D profile; We performed the ELLPROF task in BBarolo \citep{teodoro_3d_2015-1}, which uses standard tilted-ring models, and we also implemented the following simple procedure to get a profile to test for consistency.

The semi-major axis of the galaxy was determined by measuring the flux in the moment-0 map along a straight line along the position angle (PA) reported in Table 1. The line was terminated when the flux reaches four times the RMS noise in the moment map. The RMS noise in the map was determined by the noise in one channel multiplied by the square root of the number of contributing channels \citep{ianjamasimanana_smooth_2018}. 

The semi-minor axis was calculated with the same procedure repeated along a line perpendicular to the PA along the centre of the galaxy and thus an axis ratio is calculated.
A central pixel, corresponding to the chosen dynamical centre derived in Section 4.3 was chosen, and ellipses were placed on the image, their size starting with the axes lengths and then decreasing with a set value, while maintaining the axis ratio and PA of each. 
The image was broken into annular regions, with each ellipse being an annular boundary; the mean flux was calculated for each region, as was the standard error on the mean value. 
Note that the ellipses are not being fit, they are determining spatial bin regions.

    \begin{equation}
    SE= \sigma/\sqrt{n}
    \label{SE}
    \end{equation}    
In Eq.\ref{SE}, SE, $\sigma$ and n are the standard error, the standard deviation of the flux values and the number of pixels within each in the annulus, respectively. 

Thus the total uncertainty was calculated for each annulus using Eq.~ \ref{error}, with the RMS being calculated as described earlier in this section: 
    \begin{equation}
     uncertainty=\sqrt{SE^{2}+ \frac{1}{n(RMS^2)}}
     \label{error}
     \end{equation}
     
These values are then corrected for inclination presented in Table \ref{table1}. Hence, we obtain the profile seen in Fig.~\ref{profile}, which is consistent with that obtained from the ELLPROF task in BBarolo. Thus, we could proceed to the Sérsic modelling. 

The Sérsic profile is given by:
    \begin{equation}
        I(r)=I_{e}exp(-b_{n}[\frac{r}{R_s}]^{1/n}-1)
    \end{equation}
    where $b_n = 1.9992n - 0.3271$ and $I_e$ is the intensity at effective radius $R_s$ and n is the Sérsic index. 

We fitted a single-component Sérsic model and a two-component model to the derived profile and compared them with the AIC and BIC \citep{burnham_model_2010}.
Both methods provide us with a metric of comparison for fit models. However, the BIC heavily penalises large number of parameters,  and hence can provide a counter against overfitting and is great to test out a two-component fit.

While comparing models, lower AIC and BIC values are preferred. AIC for the single component Sérsic fit is 189.11 while for the two-component fit we have 57.42; AIC overwhelmingly supports the two-component fit. 
The BIC value for the single component fit is 28.65 and 22.49 for the two-component fit; in terms of probabilities given by:

\begin{equation}
    P_i = \frac{e^{-\frac{1}{2} \Delta \mathrm{BIC}_i}}{\sum_j e^{-\frac{1}{2} \Delta \mathrm{BIC}_j}}
\end{equation}

The two-component model is favoured by a factor of 20 by the BIC. Hence both criteria point to a two-component distribution of HI in WLM. Both in stars and gas, there are an inner compact component and an extended, more disc-like, outer component. The Sérsic indices for the respective components are also in agreement as shown in the legend of Fig.~\ref{profile} indicating that the two components in the gas and stars are very similar.

In Section 4 we model the galaxy using the tilted ring method. However, the model is applied separately for the approaching and receding sides owing to the galaxy's kinematic asymmetry.

\section{Dynamics of WLM}

\subsection{The asymmetric rotation curve of WLM}

\citet{kepley_high-resolution_2007} identified that WLM shows an asymmetric rotation curve on the approaching and receding sides.  We can visually inspect the moment-1 map in Fig.~\ref{mk64} to see that the blue and the red side velocities show a symmetric rapid increase in the centre, but the blue side ones taper down with sparsely packed contours while the red side keeps increasing rapidly.

The velocity contours in the moment-1 map significantly deviate from perpendicularity to the major axis, particularly in the central region. This, together with the fact that blue and red sides show different directions in the centre could be a signature of a dense bar-like structure at the centre, and we discuss this later in Section 6.

The kinematic asymmetry is also apparent if we examine the PV diagram of WLM in Fig.~\ref{PV}, where the approaching side appears to flatten after 1 kpc while the receding side continues to rise.

\begin{figure}[h!]
  \resizebox{\hsize}{!}{\includegraphics{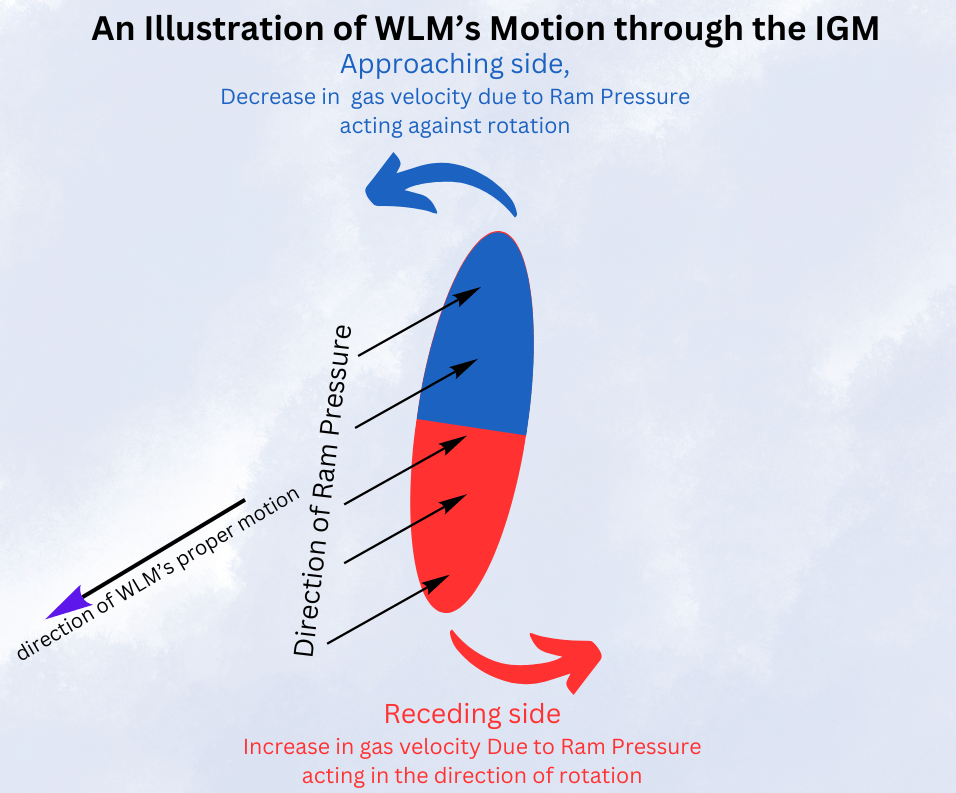}}
  \caption{\ {Illustration of how WLM's motion through the IGM might affect its kinematics. As WLM moves through the IGM, the approaching side slows as the ram pressure pushes against its rotation, while the receding side moves along the direction of ram pressure, and gains velocity, as it is seen in the rotation curves shown in Fig.~\ref{RC}}}
  \label{toymodel}
\end{figure}

\begin{figure}[h!]
  \resizebox{\hsize}{!}{\includegraphics{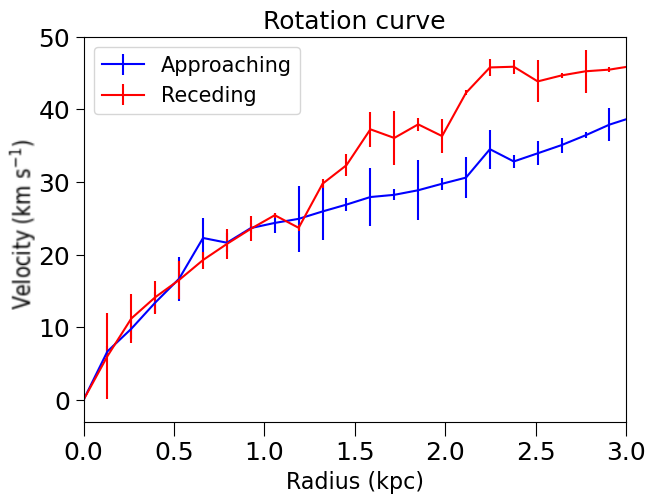}}
  \caption{\ {Asymmetric-drift-corrected rotation curve of WLM along its approaching and receding sides. The rotation remains symmetric within the inner 1.5 kpc, beyond which the curves begin to diverge, suggesting that the gas is out of equilibrium, likely due to ram pressure effects.}.}
  \label{RC}
\end{figure}

\citet{kepley_high-resolution_2007} used a 2D tilted ring model to obtain the rotation curve on both sides of the galaxy separately. Within the first kiloparsec, the approaching side of the rotation curve rises above the receding side by ~$5$ km s$^{-1}$, then the velocities cross over where the receding side rapidly rises like a solid body till the outskirts, while the approaching side shows a gentle rise. An asymmetric distribution of velocities is also apparent in the global velocity profile shown in Fig.~{\ref{globalprofile}}, it points to a multi-component kinematic structure of the galaxy. This will be discussed in detail in section 5.

\subsection{The expected impact of ram pressure on the asymmetric rotation curve }
\citet{yang_evidence_2022} used the proper motion of WLM obtained by \citet{battaglia_gaia_2022} from \textit{Gaia} DR3. 
The stripped gas is exactly in the opposite direction of the proper motion, which indicates that the ram pressure is applied in the same direction. As WLM moves through the IGM, its approaching side's motion,is aligned against the direction of ram pressure, while the receding side's motion is aligned along the direction of ram pressure. This scenario is depicted in Fig.~\ref{toymodel}.

\subsection{The final rotation curve of WLM}

To model the complex kinematics of WLM, we adopted an iterative prescription based on the 3D tilted ring fitting tool TiRIFiC \citep{tirific2007}. It allows for comprehensive disc modelling with a vast suite of available parameters to fit, with the possibility of granular control over the fitting procedure. 
We modelled the two halves of the disc separately owing to the asymmetry along the approaching and receding sides. 
All parameters were independently determined for both galaxy sides, which however share a common systemic velocity and centre.
The initial guesses for all parameters were used from \cite{ianjamasimanana_meerkat-16_2020} and \citet{kepley_high-resolution_2007}.

In the first iteration, we provided initial guesses for rotational velocity, dispersion, surface brightness, systemic velocity, PA, inclination, dynamical centre and thickness  for 14 rings on both sides of the galaxy. Only the surface brightness, systemic velocity and dynamical centre are allowed to vary and to arrive at a best fit in the first iteration. 

Next, in the second iteration, we used the previously determined parameters as initial conditions and let all of the mentioned parameters from the first iteration vary and then be determined further. Letting both the position and inclination vary all the way from the inner region to the outskirts should let us capture twist-like features in the galaxy.
Then, to model non-circular motions, we fitted for radial velocity and some second-order harmonics for velocity (RO2A, RO2P, RA2A, and RA2P in the TiRIFiC documentation) used by \citet{Spekkens_non_circular} to implement effects of bar streaming. Velocity harmonics were also used by \cite{ianjamasimanana_meerkat-16_2020} to successfully model non-circular motions in WLM.  We then compared the PV diagram and global velocity profile of the model for consistency, and then doubled the number of rings to 28 to capture smaller variations and rerun the fit with all the previous fit parameters as initial conditions.  The model was then processed through TRMerrors\footnote{\url{github.com/PeterKamphuis/TRM_errors}} to obtain errors. Acquiring errors on all the parameters for every ring is very computationally expensive. Hence, we computed errors only on rotational velocity, dispersion, inclination and PA (see Appendix \ref{tilted_param}). Values for inclination and PA presented in Table 1 are average values of all the fits presented in Appendix \ref{tilted_param}

Fig.~{\ref{globalprofile}} includes a comparison of the observed global velocity profile of WLM and of the TiRIFiC model obtained from the above procedure. The observation and the model seem to agree well.
Comparing the PV diagrams of the model and the observation in Appendix \ref{AppendPV}  , we see that all
observed features are recovered by the model, including the
asymmetry between the approaching and receding sides as well as the two-component-like split on the approaching side. 
We also provide moment maps and residuals for the TiRIFiC model in Appendix \ref{TiRiFIc_moments}.

\cite{kepley_high-resolution_2007} used a 2D method to obtain their rotation curves, and did not fit for dispersion and thickness. 
A similar prescription using TiRIFiC and velocity harmonics has been used to model WLM earlier by \cite{ianjamasimanana_meerkat-16_2020}, but they model both sides together with a single disc. 

Our approach includes independent modelling of both sides, along with a fit for thickness, dispersion and velocity harmonics, leading to a comprehensive modelling of WLM. 
These new rotation curves seen in Fig.~ \ref{RC} for the approaching and receding sides, align well within in the first kiloparsec, and do not show a divergence in the central region as it is seen in the \cite{kepley_high-resolution_2007} rotation curves. 

In Sections 5 and 6 we discuss in detail the inner component of WLM and the possibility of a bar-like structure in the centre. Such a structure can provide inner disc velocity asymmetries, and contribute to non-circular motions, which we have considered and fit for in our TiRIFiC modelling. \citet{Sofue_innerRC_MC_2025PASJ} report that for the rotation curve of the Milky Way along its inner bar region, the variation along two different regions of the Galaxy is 10-15\%. If a similar impact were to be assumed for WLM, it would be 2-3 km s$^{-1}$ at 1.5 kiloparsec for our rotation curve. The asymmetry beyond 1.5 kiloparsec that has been found by \cite{kepley_high-resolution_2007} is consistent with our results. 
 
\begin{figure*}[hb]
\centering
   \includegraphics[width=15cm]{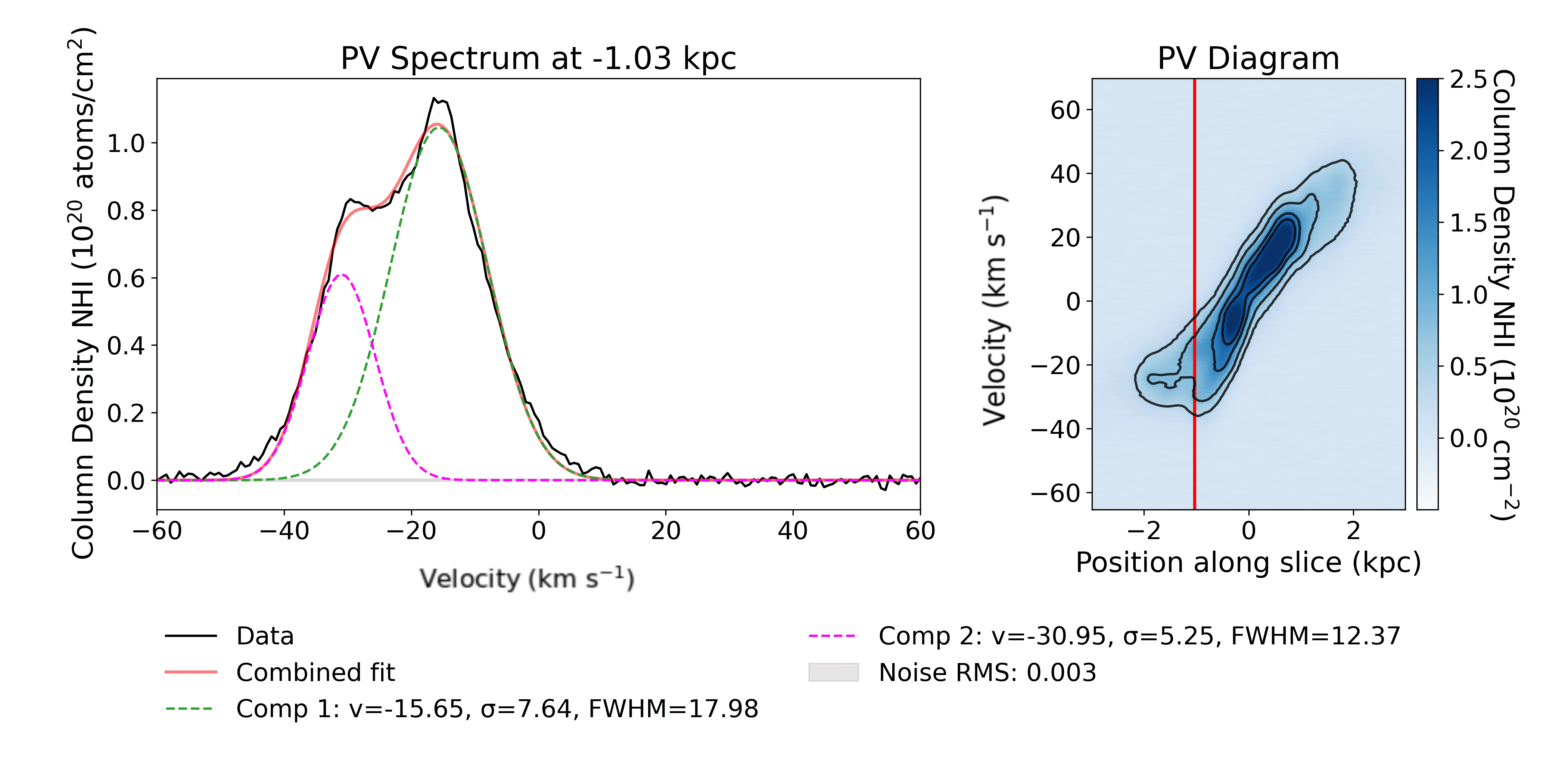}
     \caption{\ {One HI spectrum of WLM taken along the PV diagram. In the right panel, the vertical red line indicates the position in the PV diagram where the spectrum was taken. This slice shows two distinct peaks, modelled by two Gaussian fits corresponding to two different kinematic components. Parameters of the Gaussian fits are given at the bottom of the figure. Spectra around the centre and receding side of the galaxy are given in Appendix~\ref{otherslic}.}}
     \label{specdecomp}
\end{figure*}
Beyond about 1.5 kiloparsec from the galaxy centre, the restoring force is not strong enough from the galaxy centre. Then, the approaching side loses velocity and the receding side gains velocity as seen in the rotation curve's divergence shown in Fig.~\ref{RC}, and it provides further evidence that ram pressure may explain the asymmetric rotation curve of WLM (see Fig.~ \ref{toymodel}). 
We observe an increased dispersion on the eastern edge of the galaxy, which can be seen in the moment-2 map of Fig.~{\ref{mk64}}. This region is also visible in the residual moment-1 and -2 maps presented in Appendix \ref{TiRiFIc_moments}.
This is also the side of the galaxy on which ram pressure is exerted, complex turbulent gas motions and ram-pressure-induced star formation may explain this increased dispersion. 

When the rotation curve of a dwarf system does not exceed a few tens of km s$^{-1}$, it becomes essential to apply the asymmetric-drift correction. More specifically, for gas discs with ratios of rotational velocity to velocity dispersion as low as Vrot/$\sigma_v<$ 4, pressure support can significantly affect the kinematics and must be taken into account using the so-called asymmetric-drift correction \citep{binney2011galactic,Bureau_Carignan_2002,Oh_2011}. We implemented the procedure described by \cite{littlethings_mass}. However, this correction is only necessary to estimate the dynamical mass of the galaxy under the assumption of equilibrium, which will be examined in a future paper. Corrected rotation curves are shown in Fig.~ \ref{RC}, uncorrected ones are presented in Appendix \ref{tilted_param}.

\subsection{The shift in the gas dynamical centre due to ram pressure}

As described earlier, we let TiRIFiC find the dynamical centre and systemic velocity in our iterative fitting procedure, common for all rings on either side of the galaxy, and it appears to be shifted by about 90 parsec from the stellar centre in the direction of the ram pressure being applied to the galaxy, as reported in table \ref{table1}. Further, we would like to note that the shift is within 1 beamsize and could be impacted by smearing.

Ram pressure is expected to affect the dynamics of the low-column-density gas-disk outskirts more, while gas in the inner region of the galaxy is denser and has experienced more orbits. For instance, looking at the PV diagram in Fig.~\ref{specdecomp} we can see that the rotational velocity is roughly 20 km s$^{-1}$ at 1 kpc, and we can estimate the number of orbits the gas has had at that radius to be 6.6 in 2 Gyr, while taking a rough velocity of 30 km s$^{-1}$ at 2.5 kpc we get $\sim$ 4 orbits in 2 Gyr.
Therefore, we expect the inner region to be more at dynamical equilibrium than the outer one, and we do confirm that the approaching and receding sides coincide in the inner region (see a discussion of galactic disc dynamics in \citealt{hammer_dark_2025}). 
It is not uncommon for dwarf irregulars to have their HI centres slightly different from their optical centres derived from photometry \citep{patra_figgs2_2016}. 
WLM seems substantially perturbed by ram pressure, with having lost 13\% of its HI mass, just in the MeerKAT-16 field of view, and this interaction may contribute to the shift of the dynamical centre.

\section{Kinematic two-component decomposition}

\begin{figure*}[!htbp]
\centering
   \includegraphics[width=17cm]{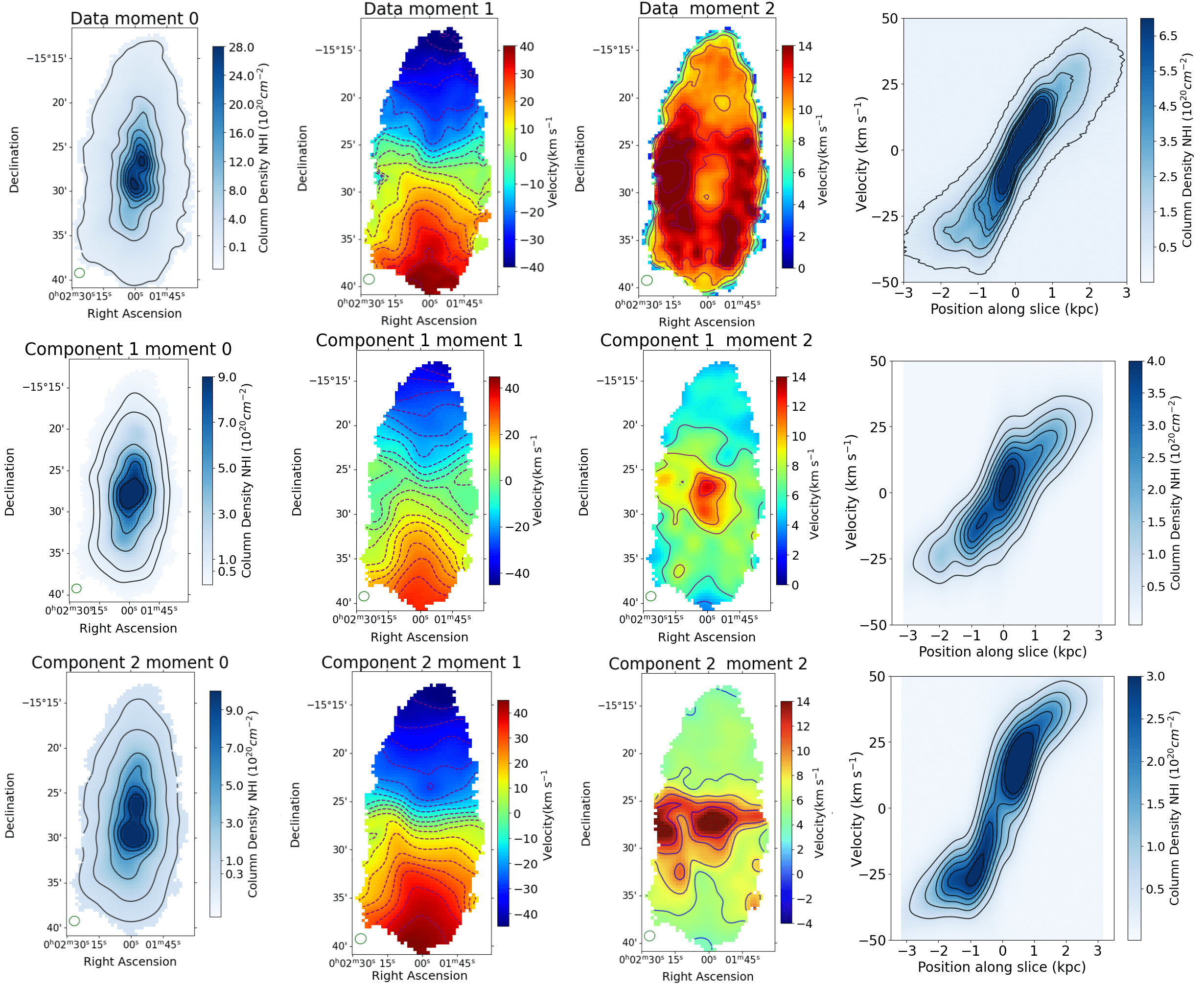}
     \caption{First row shows the moment maps from our MeerKAT observations. Second row shows the moment maps of the slower component from our decomposition, which is also the central compact component. Row 3 has maps for the faster component, which is the more extended disc-like component. Each row is presented with a corresponding PV diagram to the right. The contours on the moment-0 map and the PV diagram correspond to the ticks on the colour bar. The contours on the moment-1 map correspond to a velocity range between -40 and 40 km s$^{-1}$ with a spacing if 5 km s$^{-1}$ and for the moment-2 map they correspond to 6,9 and 12 km s$^{-1}$, respectively. }
     \label{2comp}
\end{figure*}
\begin{figure*}[!htbp]
\centering
   \includegraphics[width=15cm]{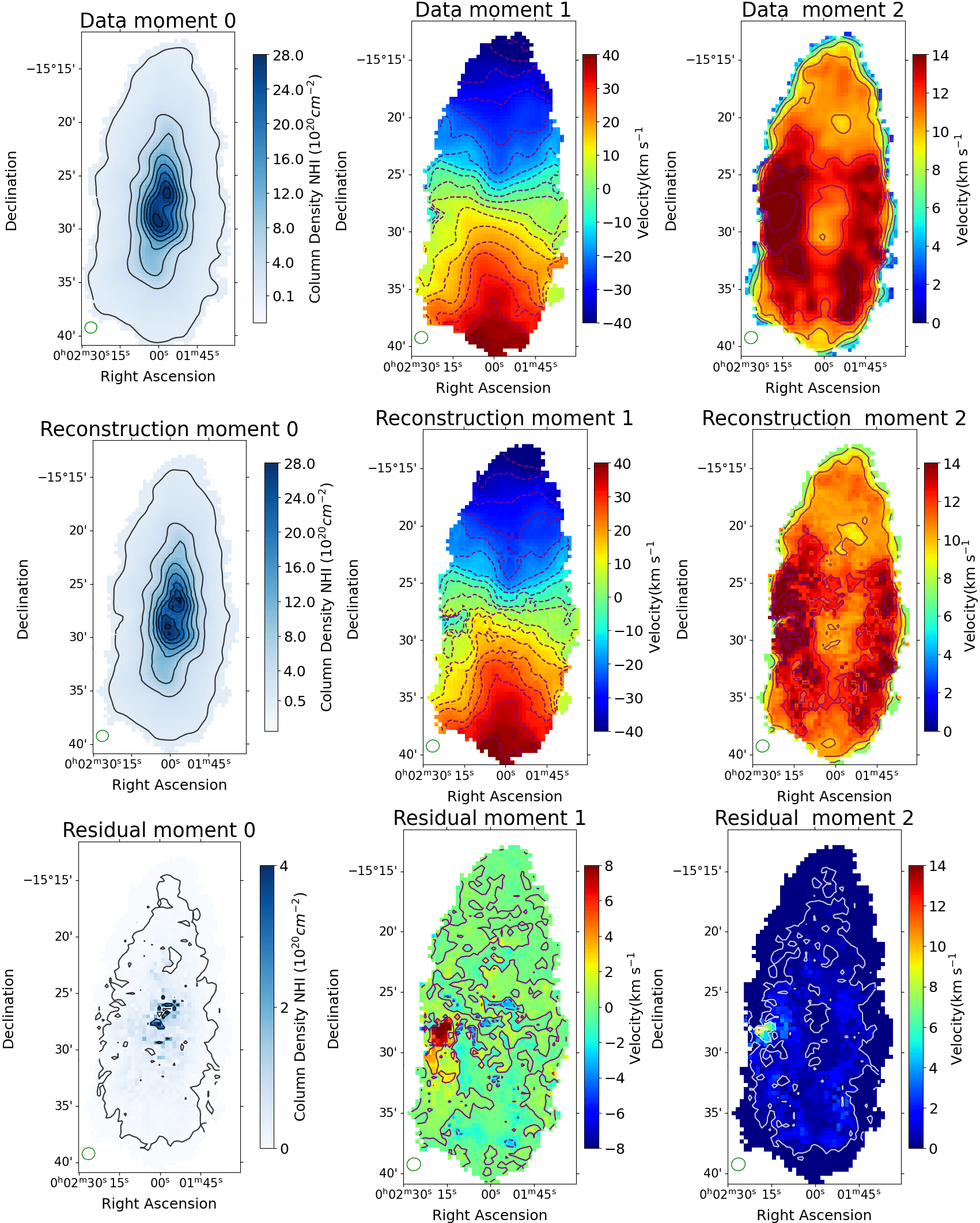}
     \caption{\ {The first row shows moment maps of the Meerkat-64 observations, the second row shows our combined two-component model which reproduces most of the features in the observed moment maps, the bottom row shows the residuals from the subtraction of the two. The contours of moment-0 map correspond to the ticks of the colourbar. The contours on the moment-1 map correspond to a velocity range between -40 and 40 km s$^{-1}$ with a spacing of 5 km s$^{-1}$ and for the moment-2 map they correspond to 6,8,10,12 and 14 km s$^{-1}$, respectively.   }}
     \label{2compcomb}
\end{figure*}

 The global flux profile we present in Fig.~\ref{globalprofile} shows a strong peak at 10 km s$^{-1}$  and a plateau-like feature between 0 and -25 km s$^{-1}$. This suggests that there could be more than one rotating components in the gas body of WLM. 
 Also, in the PV diagram presented in Fig.\ref{specdecomp}, on the approaching side we can see that the flux distribution also suggests the presence of two distinct kinematic branches.
 Further evidence of more than one component could be seen if we observe the spectrum taken in the region beyond 1 kpc as seen in Fig.~\ref{specdecomp}, which shows two distinct peaks, and indeed a two-component Gaussian fits much better to the spectrum in this region. On the receding side (Appendix~\ref{otherslic}, row 3), the second component may not be resolved, but the spectrum here is much broader than it is at the centre of the PV diagram and is also better modelled by two components.

To systematically decompose our data cube into these two distinct components, we have applied the above method to every pixel in the data cube instead of just the central slice of the PV diagram. 
First, the original data cube has been re-binned by a factor of 4:1 to increase the signal-to-noise ratio.  We then extracted the spectrum at every pixel in the data cube and fitted two Gaussian profiles to the spectrum in each pixel in the data cube akin to the fit seen in Fig.~\ref{specdecomp} (see also Appendix \ref{otherslic}).

We segregated the components into two separate data cubes; at each pixel, we sorted the component with the lower velocity of the two with respect to the systemic velocity into what we refer to as Component-1 and the larger velocity into what is referred to as Component-2. A beamsize smoothing kernel is applied to compensate for spurious fits from pixel to pixel.
Then, we generated moment maps for these two components (see Fig.~\ref{2comp}).

\begin{figure}[!htbp]
  \resizebox{\hsize}{!}{\includegraphics{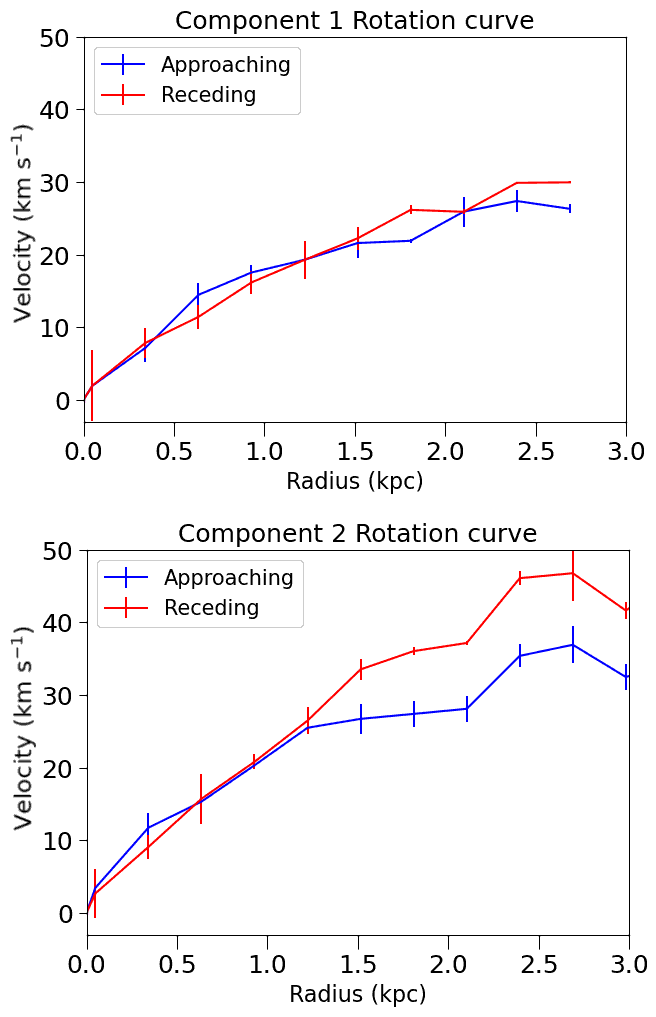}}
     \caption{\ {Rotation curves of each individual component. Component-1 shows a solid-body-like rotation curve that seems unaffected by ram pressure. The disc-like Component-2 shows a strong divergence beyond 1 kpc and is likely perturbed by ram pressure.   }}
     \label{2comprc}
\end{figure}

The reproduction of the datacube after this two-component decomposition can be seen in the second row of Fig.~{\ref{2compcomb}}. We can see that it near perfectly reproduces the column density distribution of HI in the moment-0 map, as well as the velocity field in the moment-1 map. There are minor local variations from the observations in the moment-2 dispersion map, but the global features are well reproduced. 
 We also computed rotation curves for each component using the methodology we used for the global rotation curve (see Section 4 and Fig.~ \ref{2comprc}).
 
\begin{figure}[!htbp]
  \resizebox{\hsize}{!}{\includegraphics{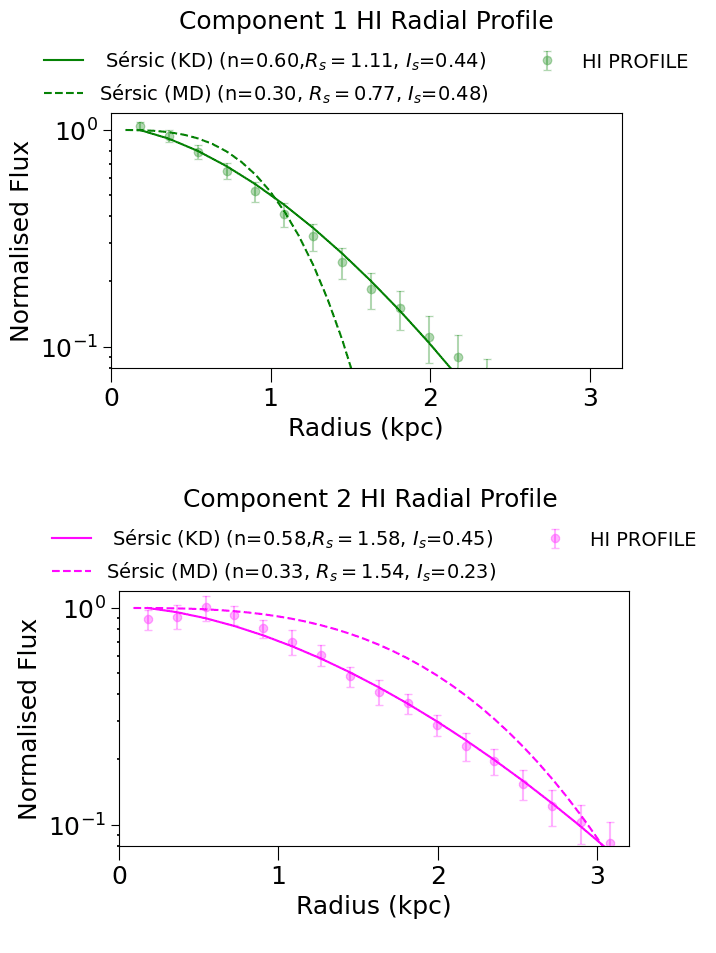}}
  \caption{\ { 1D HI profiles of the kinematically decomposed (KD) components fit with a Sérsic profile. The Sérsic parameters are listed at the top of each panel. We have over-plotted with dashed lines the Sérsic profile from our morphological decomposition (MD) from section 3.   }.}
  \label{newprof}
\end{figure}

In Section 3, we obtained a two-component Sérsic fit on the total HI profile of WLM. This let us establish that the two components present in the stellar profile of WLM reported in \cite{higgs_solo_2021} can be mapped onto the HI profile. 
Due to the high inclination of the disc, these components overlap in the global moment-0 map of WLM. The kinematic decomposition of the data let us separate the two components cleanly and obtain the moment maps presented in Fig.~{\ref{2comp}}.
We could then fit a single Sérsic profile individually to the two kinematically separated components and compare them to the Sérsic fits from Section 3. This comparison can be seen in Fig.~{\ref{newprof}}.
Both the kinematic decomposition and morphological decomposition show a compact inner component and a sparse outer component. 
Though in both cases the distinction between a compact and a sparse component is very clear, the Sérsic parameters do not match perfectly, which is expected due to projection effects and the overlap between the two components. We discuss the properties of the two components in detail in Section 6.

\begin{figure}[!htbp]
  \resizebox{\hsize}{!}{\includegraphics{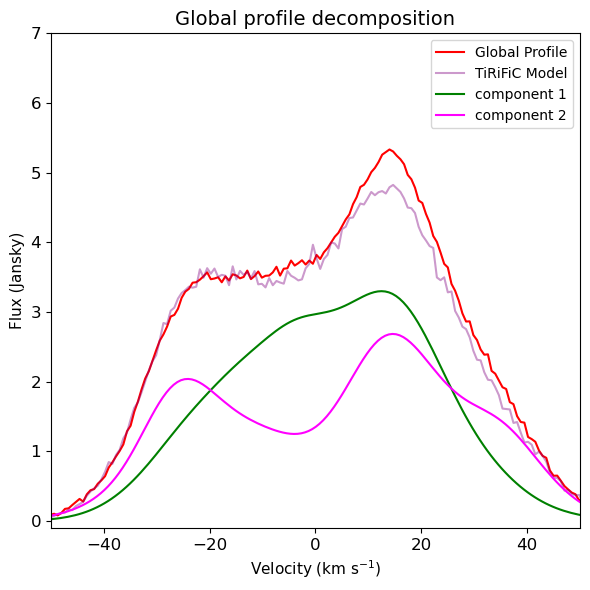}}
  \caption { Emission profile of the whole body of WLM with the contributions from the two components along with the profile of the TiRIFiC model.}
  \label{globalprofiledecomp}
\end{figure}
We also obtained a global velocity profile for the two components that we present in Fig.~{\ref{globalprofiledecomp}}. Component-1 spans a smaller velocity range but contributes more to the sharp peak in the profile. This can be verified from the rotation curve, the velocity amplitude of this component is lower at all radii than that of Component-2. Component-2 is much more widely distributed in the velocity space, extends beyond Component-1, and shows the  differential rotation shape expected for a rotating disc, i.e. a double-horn profile. Fig.~{\ref{globalprofiledecomp}} shows that Component-2 has a slight velocity excess of few km s$^{-1}$ on the receding side at about 40 km s$^{-1}$, which might be due to ram pressure.

\section{Discussion}

\subsection{Inner Component-1}

We now refer to Component-1 as the inner component of WLM. It has a half-light radius of 0.77 kpc in the morphological global profile decomposition presented in Section 3 and 1.11 kpc in the profile obtained after the kinematic decomposition in Section 5 and presented in Fig{\ref{newprof}}. For comparison, Component-2 is the outer component with a half-light radius of 1.53 kpc and 1.58 kpc in the global profile and kinematic decomposition results, respectively.

Gas in the moment-0 map for this component is more centrally packed, along both the major and minor axes. The central iso-velocity contours in the moment-1 map of Component-1 are oriented along a PA of about 165 degrees ~ as seen in  the second row of Fig.~{\ref{2compcomb}}, similar to the moment-1 map from the non-decomposed cube in the first row. In contrast, the inner velocity contours of Component-2 are nearly perpendicular to the PA of WLM.
This suggests that we have successfully separated the component that contributes to the  165 degrees contours in the global moment-1 map.
This off-axis orientation also points towards this component's bar-like nature.
The PV diagram for this component does not show any flattening at the outskirts and shows solid-body like increase in rotational velocity on both the approaching and receding sides all the way till at least 1.7 kpc. However, a careful examination of the PV diagram indicates a very central component, whose orientation is shifted from the overall PV diagram. This suggests component-1 to be more complex than a single bar. We will study the impact of such a structure on the dynamics of a galaxy like WLM interacting with the IGM in a future study with full hydrodynamical modelling. This is also seen in the rotation curve of this component. These bar-like motions contribute to the steep rotation curve in the inner regions. WLM is well known in the literature for a high star formation rate at the centre of the galaxy \citep{albers_star_2019}. Along with the high inclination of WLM, this renders classical tests such as the Tremaine-Weinberg method for bars on a gas velocity field inconclusive \citep{hernandez_relevance_2005}. Gaseous bar-like structures have also been reported in several dwarf irregulars such as NGC 6822, NGC 3741, NGC 2915 and DDO 168 \citep{NGC_2915_Meurer,banerjee_slow_2013,patra_detection_2019-1}.

 \citet{g_are_2025} have also recently shown that gaseous bars can form and remain stable through hydrodynamical simulations.
Another important feature that points towards a central bar-like structure is the distinct two-component split seen in the PV diagram. \citet{1999AJ....118..126B} show that a separate kinematic component can be a diagnostic for boxy peanut-shaped structures in galaxies.
An ancillary support to the component to be a bar-like also comes from its Sérsic index of 0.6, as \citet{10.1007/978-3-540-75826-6_15} suggested that bars usually have Sérsic indices between 0.5 and 1.
 
\subsection{Outer Component-2}

The outer Component-2 in both the morphological and kinematic decomposition shows extended and disc-like features. 
It seems to trace a more spread-out component of the gas distribution along the major and minor axes of the galaxy. Looking at the velocity field we see that at inner radii, we have a rapid increase in velocity on both sides of the galaxy, which progressively evolve into V-shaped contours often seen in galactic discs. The central iso-velocity contours in the moment-1 map are nearly perpendicular to the PA= 175 degrees major axis, indicating clear distinction from the off-axis bar-like Component-1.
The dispersion of this component decreases progressively from the centre to the outskirts; such steady variation is often seen in galactic discs supporting the validity of our decomposition \citep{doi:10.1142/q0016}. The classical double-horn profile, which is the kinematic signature of HI rotating discs is seen in Component-2 is seen in Fig.~\ref{globalprofiledecomp}. The two horns correspond to the outer parts of the disc where the rotation curve is flat, producing a large contribution at the maximum projected velocities. The relative dip near the systemic velocity arises because gas at central velocities comes from smaller radii and contributes less to the total flux.  In addition, the double-horn shape becomes more distinct and stronger as galaxies are viewed closer to edge-on where the observed line-of-sight velocities closely reflect the true rotational velocities \citep{Giovanelli_Haynes_1988}.
In the PV diagram for this component in Fig.~\ref{2comp} we can clearly see velocities flattening beyond one kiloparsec on the approaching side, which is also seen in its rotation curve  in Fig.~\ref{2comprc}.

Component-2 is sparse at the outer edges, and it seems to have been perturbed by ram pressure, which appears to be responsible of the rotation curve asymmetry observed in the global rotation curve.

\section{Conclusions}

WLM is a gas-rich LG dwarf irregular whose gas kinematics and dynamics are perturbed by ram pressure effects exerted by the surrounding IGM. \citet{yang_evidence_2022} show that WLM is continuously losing gas as it travels through the IGM.
Our revised MeerKAT-16 moment maps show that 13 \% of the total gas mass has been lost from the main body just in the field of view.  Considering that WLM is a gas-rich and star-forming dwarf galaxy at the current epoch, its dynamics was likely dominated by gas motions in the past. Using our new high-resolution MeerKAT-64  observations,  we have been able to show that the gas kinematics of WLM beyond 1.5 kiloparsec is severely impacted by its motion through the IGM, which consequently affects the rotation curve and any subsequent mass estimates based on it.

Our morphological and kinematic decomposition shows that
WLM appears to have an inner bar-like structure that is identified by its solid-body inner rotation curve; this structure seems more resilient to perturbations. The disc-like outer component is characterised by a double-horn profile and is severely impacted by ram pressure effects, which contribute to the rotation curve asymmetry between the receding and the approaching sides.

Dynamical studies of dwarf galaxies often focus on isolated bodies with minimal tidal interactions to better understand their internal structure and in turn estimate their total mass. Here we find that dwarf motions with respect to the local IGM can generate sufficiently strong ram pressure effects to affect the gas dynamics at the outskirts of their disks. It suggests that more consideration must be given to understanding the distribution and density of the ionised medium through which galaxies travel. 

In the near future, we will present detailed hydrodynamical simulations in an attempt to model WLM and its interaction with the IGM to further probe the above scenario and further probe how ram pressure affects the galaxy differentially.
The LG IGM is sparse enough to be undetected by present-day direct observations. 
Accurately modelling WLM and its interaction with the medium will allow us to place realistic constraints on the local IGM density as well as on the WLM's total mass.

\begin{acknowledgements}
Authors RI and BN acknowledge financial support from the grant PID2021-123930OB-C21 funded by MICIU/AEI/ 10.13039/501100011033 and by ERDF/EU, and the grant CEX2021-001131-S funded by MICIU/AEI/ 10.13039/501100011033, and the grant TED2021-130231B-I00 funded by MICIU/AEI/ 10.13039/501100011033 and by the European Union NextGenerationEU/PRTR, and the grant INFRA24023 (CSIC4SKA) funded by CSIC, and acknowledge the Spanish Prototype of an SRC (espSRC) service and support funded by the Ministerio de Ciencia, Innovacion y Universidades (MICIU), by the Junta de Andalucia, by the European Regional Development Funds (ERDF) and by the European Union NextGenerationEU/PRTR. The espSRC acknowledges financial support from the Agencia Estatal de Investigacion (AEI) through the Center of Excellence Severo Ochoa award to the Instituto de Astrofisica de Andalucia (IAA-CSIC) (SEV-2017-0709) and from the grant CEX2021-001131-S funded by MICIU/AEI/ 10.13039/501100011033. Part of BN work was supported by the grant PTA2023-023268-I funded by MICIU/AEI/ 10.13039/501100011033 and by ESF+. The MeerKAT telescope is operated by the South African Radio Astronomy Observatory, which is a facility of the National Research Foundation, an agency of the Department of Science and Innovation.   
\end{acknowledgements}

\bibliography{final_wlm1.bib}

\begin{appendix}  

\section{Radial variation in tilted rings fit parameters}\label{tilted_param}
\begin{center}
  \includegraphics[width=15cm]{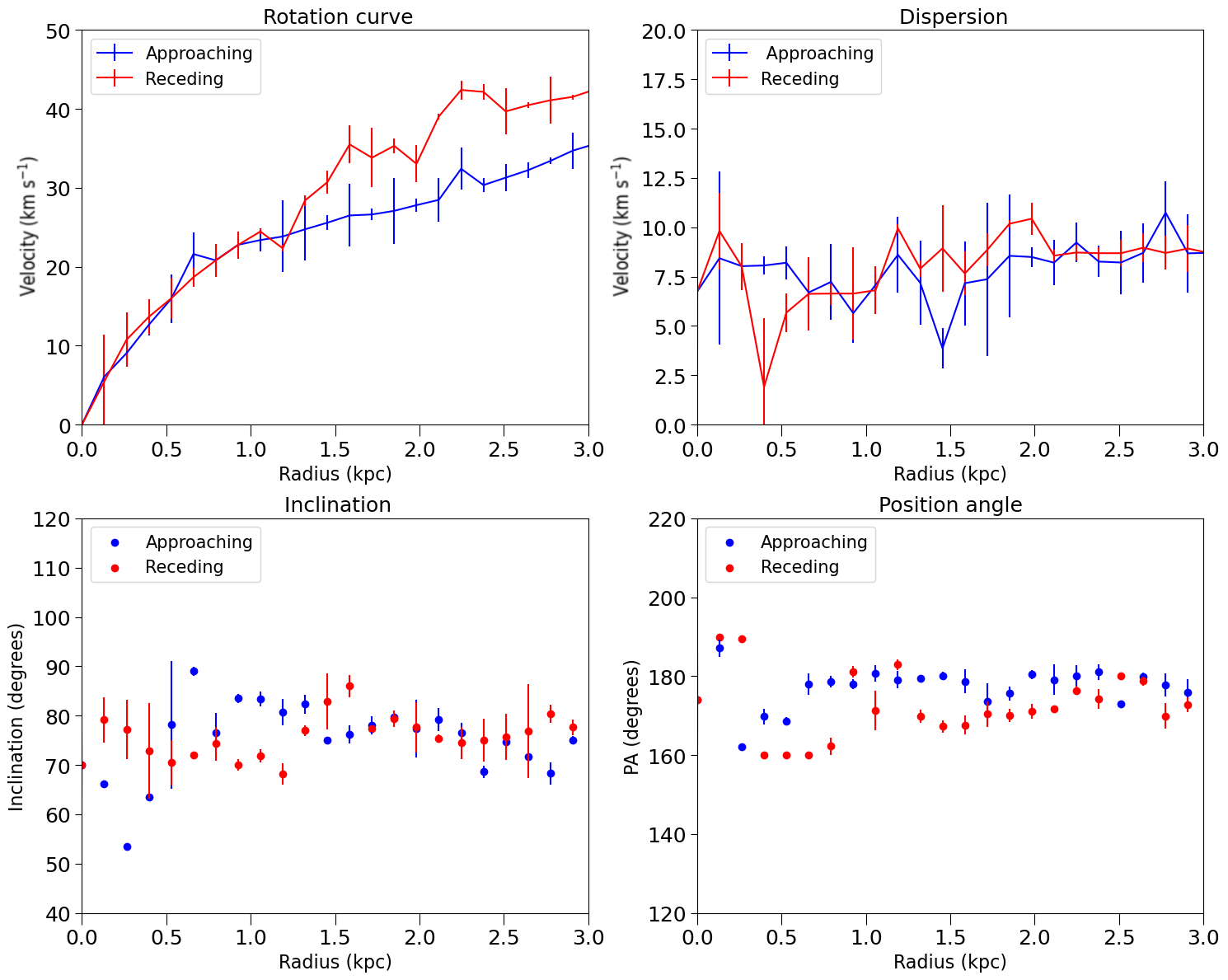} 
  \captionof{figure}{Subplots show the variation in the fit Rotation velocity, dispersion, Inclination and position angle of the tilted rings fit described in Section 4. The velocities here are not corrected for asymmetric drift}

\end{center}

\section{Comparing PV diagrams: Observations Vs Model}\label{AppendPV}
\begin{center}
  \includegraphics[width=15cm]{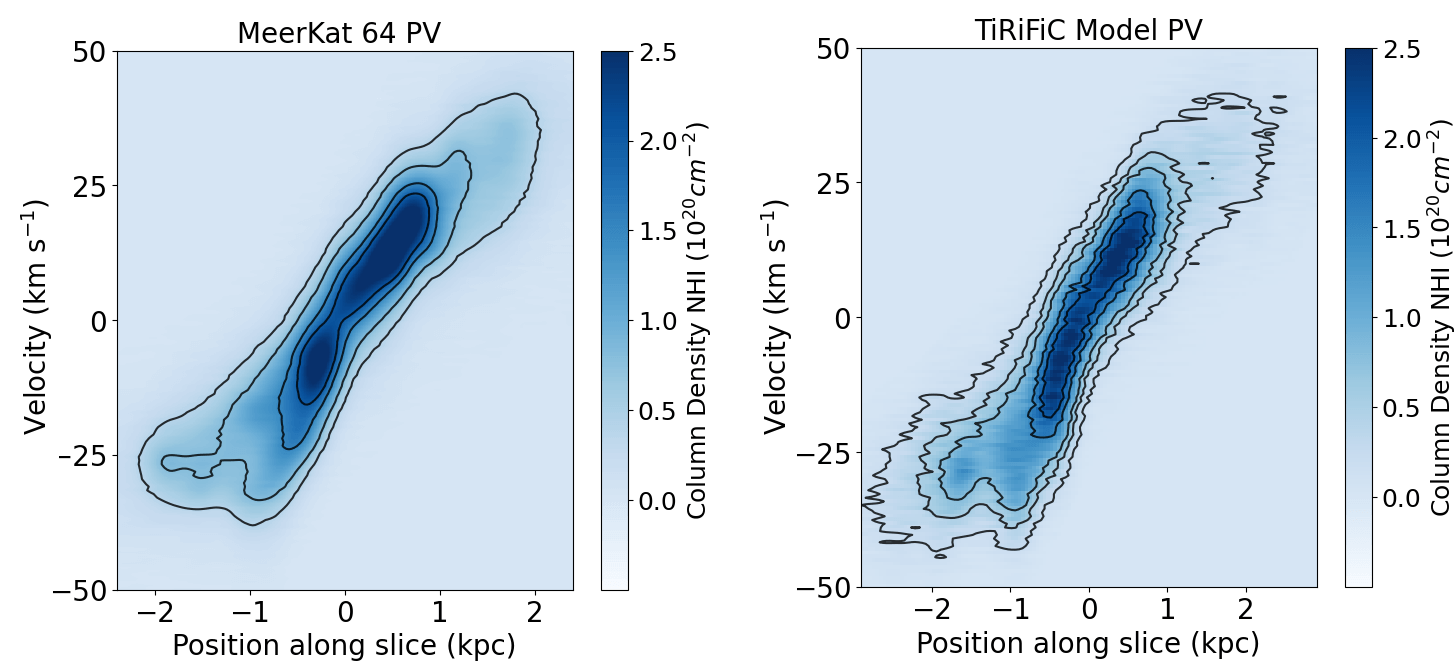} 
  \captionof{figure}{A comparison of PV diagrams from the MeerKAT observations on the left and the TiRiFiC model on the right. Contours correspond to ticks on the colourbar}

\end{center}
\clearpage

\section{TiRiFiC Model Moment Maps and Residuals}\label{TiRiFIc_moments}
\begin{center}
  \includegraphics[width=16cm]{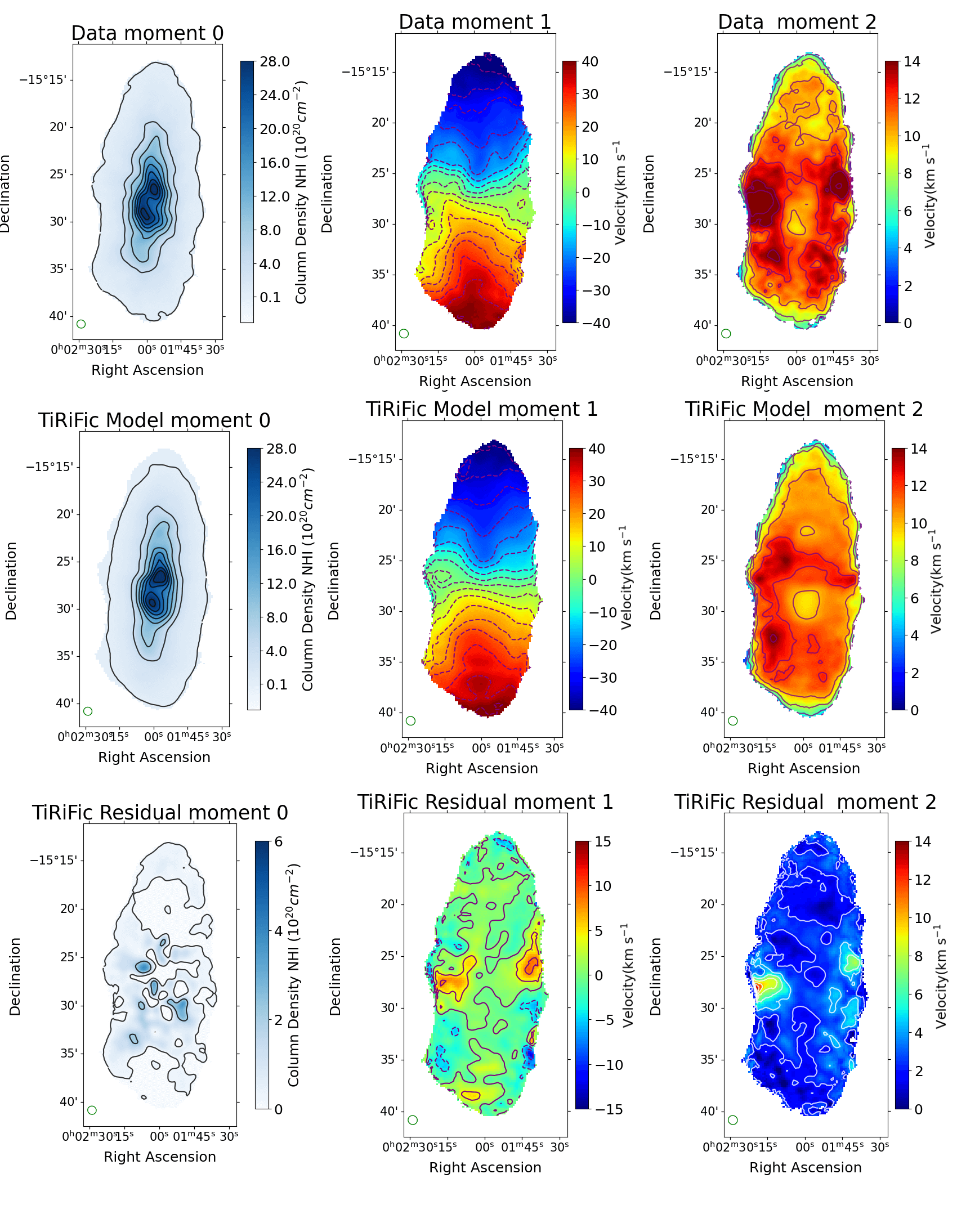} 
  \captionof{figure}{Contours in the first 2 rows and all moment 0 maps follow the same structure as described in Fig.~{\ref{2compcomb}}. For the residuals in row 3 the contours on the moment 1 map correspond to $-12$ to $12$ km s$^{-1}$ with an increment of 4 km s$^{-1}$ and for the moment 2 map in the same row they correspond to the ticks on the colourbar }

\end{center}
\clearpage
\section{Other slices of the PV diagram}\label{otherslic}
\begin{center}
  \includegraphics[width=15cm]{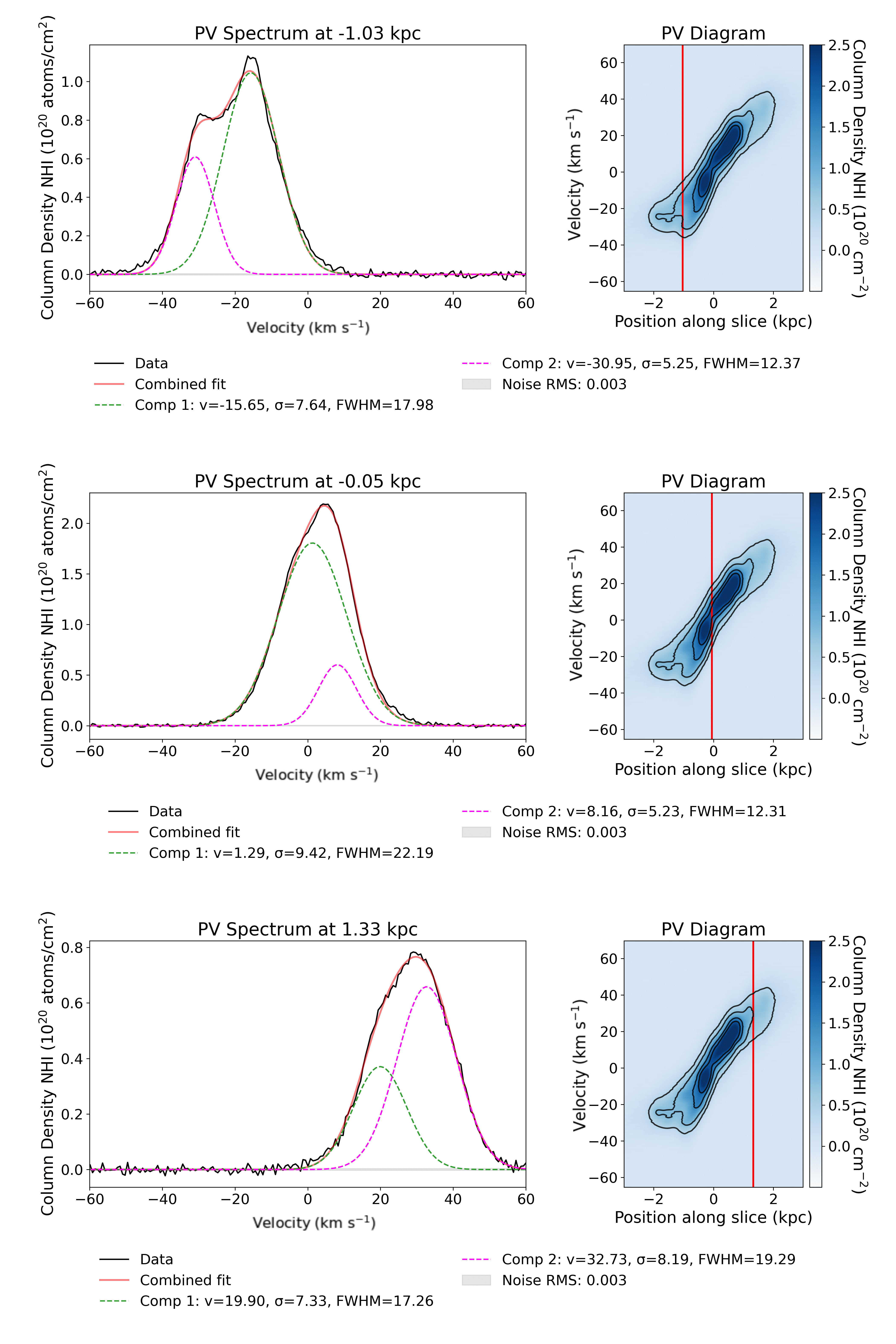} 
  \captionof{figure}{Spectra along 3 different slices of the PV diagram along with two-component decomposition. Same symbols as in Fig.~ 8.}

\end{center}

\end{appendix}
\end{document}